%% file: main.tex
\documentclass[10pt, conference, letterpaper]{IEEEtran}
\usepackage{cite}
\usepackage{amsmath,amssymb,amsfonts,amsthm}
\usepackage{algorithmic}
\usepackage{graphicx}
\usepackage{textcomp}

\usepackage{xcolor}
\usepackage{xspace}
\usepackage{float}
\usepackage{booktabs}
\usepackage[caption=false,font=footnotesize]{subfig}  
\def\BibTeX{{\rm B\kern-.05em{\sc i\kern-.025em b}\kern-.08em
    T\kern-.1667em\lower.7ex\hbox{E}\kern-.125emX}}
\begin{document}

\include{macros}

\title{Scaling the Lightning Network\\with Practical Set Reconciliation\footnote{A version of this article appeared in IEEE ICBC 2026.}}

\author{\IEEEauthorblockN{Xingyu Chen}
\IEEEauthorblockA{\textit{Boston University} \\
Boston, USA \\
chxy517@bu.edu}
\and
\IEEEauthorblockN{Anish Sinha}
\IEEEauthorblockA{\textit{Red Hat} \\
Boston, USA \\
anishs@bu.edu}
\and
\IEEEauthorblockN{David Starobinski}
\IEEEauthorblockA{\textit{Boston University} \\
Boston, USA \\
staro@bu.edu}
\and
\IEEEauthorblockN{Ari Trachtenberg}
\IEEEauthorblockA{\textit{Boston University} \\
Boston, USA \\
trachten@bu.edu}
}

\maketitle

\begin{abstract}
The Lightning Network (LN) utilizes gossip to share network topology, channel announcements and updates, and node announcements among its local constituents.  Yet, our measurements show that this flooding-based gossip reconciliation is fundamentally inefficient.  We propose, instead, to use set reconciliation protocols for sharing this information, and we systematically evaluate existing approaches under realistic network conditions. We further propose \adaptiveiblt, a novel adaptive IBLT (Invertible Bloom Lookup Table) protocol with a partial-decoding enhancement. By simulating reconciliation in Core-Lightning and evaluating real gossip snapshots, we demonstrate the practical benefits of reconciliation in scaling gossip reconciliation from hours down to a few minutes.

\end{abstract}


\section{Introduction}

The Lightning Network (LN) is a prominent off-chain scaling solution for Bitcoin and other cryptocurrencies, designed to support fast and low-cost payments by executing transactions over bidirectional off-chain (multi-hop) payment channels rather than on the more cumbersome blockchain~\cite{masteringLN}. To successfully compute payment routes, each node must maintain an up-to-date view of the network's channel graph, including channel announcements / updates, and node announcements.

Each LN node maintains a local database, commonly referred to as the \texttt{gossip\_store}, which stores all known channels and node information. When a node connects to new peers or rejoins the network after downtime, it synchronizes its gossip state by receiving gossip messages from its peers. Today, this reconciliation process relies almost exclusively on flooding: peers repeatedly exchange gossip messages regardless of how much information is already common.

Several prior studies and engineering efforts~\cite{gögge2022routingdelayLN, harvey2025gossipobserver, he2019improvedgossip} have observed that LN gossip converges slowly and consumes substantial bandwidth. Recent discussions among Lightning Network developers further highlight that these inefficiencies stem from deeper structural limitations of the current gossip protocol~\cite{lnsummit2024_gossip}. In particular, the code exhibits inconsistent synchronization, lack mechanisms to reliably detect stale or missing updates, and often rely on heuristic policies (\eg peer selection, rate limiting, or periodic resynchronization). These limitations can lead to incomplete network views (\eg failure to resurrect pruned channels) and unnecessary transmission of out-of-order or redundant messages. Such observations suggest that the inefficiency of LN gossip is not merely due to suboptimal broadcast strategies, but rather arises from the fundamental mismatch between broadcast-based dissemination and the underlying goal of state reconciliation across peers.

By contrast, we argue that the LN gossip mechanism is more naturally viewed as a \emph{set reconciliation} problem. At any point in time, two LN peers maintain highly overlapping gossip datasets and only exchange their differences. From this perspective, repeated gossip flooding is unnecessary: peers can reconcile their local states directly by exchanging compact summaries that encode only set differences.

This problem naturally maps to the classical set reconciliation problem~\cite{minsky2003set}, in which two parties seek to compute the symmetric difference of their datasets while minimizing communication and computation. A rich literature proposes reconciliation protocols based on compact sketches—such as Bloom filters~\cite{bloom70spacetime}, characteristic polynomial interpolation (CPI)~\cite{minsky2003set}, and invertible Bloom lookup tables (IBLTs)~\cite{goodrich2011invertible}. However, much of this work assumes that the size of the set difference is known in advance, or else evaluates protocols under idealized network conditions.

In practice, neither assumption holds for LN gossip. The difference between peers’ gossip states is highly dynamic and difficult to estimate, and reconciliation is performed over real TCP connections where latency, bandwidth, and round complexity strongly affect end-to-end performance. Although recent reconciliation protocols have saught to remove the need to know the set difference in advance, there has been little systematic evaluation of how such protocols behave or how they compare against one another in practical settings.

\subsection{Contributions}

Our work addresses these gaps through a combination of system-level measurement, protocol design, and transport-aware evaluation. We make the following contributions:

\begin{itemize}
    \item We perform packet-level measurements of the Lightning Network gossip protocol, showing significant redundancy and latency from its flooding-based gossip.
    
    \item We integrate the open-source GenSync~\cite{bovskov2022gensync} reconciliation framework into Core Lightning to conduct a unified, fair comparison of multiple reconciliation protocols.
    
    \item We propose \adaptiveiblt, a dynamic, runtime-adaptive, reconciliation protocol designed for unknown difference sizes and challenging network conditions.
\end{itemize}

Together, these contributions provide a practical understanding of how reconciliation can fundamentally improve the efficiency of gossip dissemination in the Lightning Network.

\subsection{Roadmap}
The remainder of this paper is organized as follows.
Section~\ref{sec:background} provides background on Lightning Network gossip and surveys prior data reconciliation protocols. 
Section~\ref{sec:ln-observation} presents an empirical study of Lightning Network gossip behavior, quantifying the communication overhead incurred by flooding-based dissemination.
Section~\ref{sec:adaptive} presents our design of \adaptiveiblt, including 
an enhancement via partial decoding.
Section~\ref{sec:evaluation} evaluates reconciliation protocols within a unified GenSync benchmarking environment.
Section~\ref{sec:ln-impl} discusses our integration of set reconciliation into the Lightning Network and reports end-to-end measurements on gossip snapshots.
Finally, Section~\ref{sec:conclusion} summarizes our key findings.

\section{Background}
\label{sec:background}

\subsection{Lightning Network Gossip Protocol}
\begin{figure}[t]
    \includegraphics[width=0.97\linewidth]{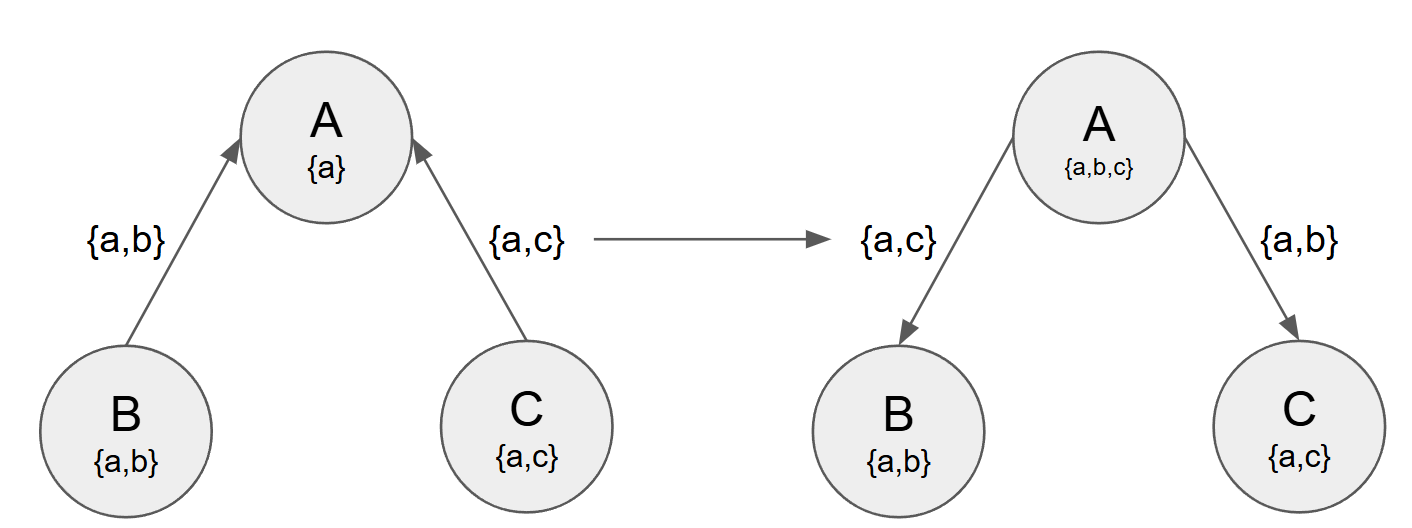}
    \caption{Flooding-based Gossip Dissemination in the Lightning Network.  Nodes maintain sets and receive update gossips from peers. Here, peers redundantly forwards $\{a\}$, though they all already it.}
    \label{fig:gossip-flow}
\end{figure}

Gossip dissemination in LN follows a flooding-based design. When a node establishes a connection with a peer, it exchanges gossip messages and forwards newly received gossip to other peers. For example, when node~A connects to node~B and node C with partial data set (see Figure~\ref{fig:gossip-flow}), it may receive gossip items that are already within its own store (e.g., $\{a\}$). To avoid infinite rebroadcasting, LN nodes track short-term message history using gossip timestamps and rate limits, but do not maintain an explicit representation of which peers already share common state.

In practice, when a node restarts or reconnects after downtime, it may need to replay a substantial amount of historical gossip to reconstruct a routing-ready view of the network. These inefficiencies have motivated engineering solutions such as Rapid Gossip Sync (RGS)~\cite{ldkRGS}, which distributes compact snapshots of the routing state to accelerate initial reconciliation. RGS relies on trusted snapshot providers and does not address the more general problem of peer-to-peer reconciliation under unknown differences.

Overall, LN has limited control over efficiency when peers possess highly overlapping routing state. For this reason, we explore alternative reconciliation mechanisms in this work.

\begin{figure}[t]
    \includegraphics[width=0.97\linewidth]{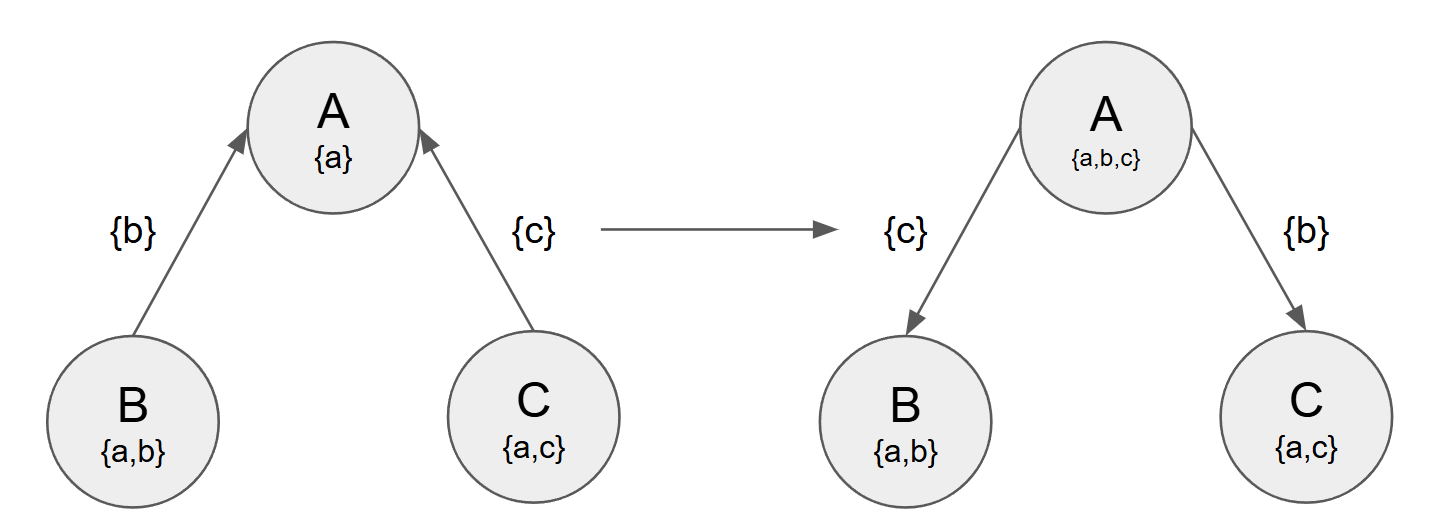}
    \caption{Example of an efficient set reconciliation workflow, that only communicates elements missing from other nodes.}
    \label{fig:sync-flow}
\end{figure}

\subsection{Data Reconciliation Protocols}
\label{subsec:protocols}
Data reconciliation enables the efficient computation of $(S_A \setminus S_B)\cup(S_B \setminus S_A)$ for two remote sets $S_A$ and $S_B$ with minimal communication. 
Figure~\ref{fig:sync-flow} contrasts the gossip workflow with a set reconciliation–based approach. 
Instead of repeated broadcasts, nodes treat local state as sets and propagate only differences. In the illustrated example (Figure~\ref{fig:sync-flow}), node~A reconciles separately
with node~B and node~C to obtain missing elements, and then propagates only newly learned items. Reconciliation directly targets state divergence instead of relying on repeated broadcast.

Classical reconciliation protocols roughly fall into two types: algebraic and filter-based. Algebraic approaches, such as Characteristic Polynomial Interpolation (CPI)~\cite{minsky2003set} and its variants \interactive~\cite{minsky2002scalable,TMC03}, encode set differences using mathematical structures, incurring low communication cost but high computation complexity. Filter-based approaches trade accuracy or robustness for efficiency.  For example, reconciliations based on Bloom Filters~\cite{bloom70spacetime} admit false positives, and those based on Invertible Bloom Lookup Tables (IBLT)~\cite{goodrich2011invertible} require accurate estimates of set difference $d$.

Several adaptive and rateless designs have been proposed to remove the need for estimating $d$. \metiblt~\cite{lazaro2023metiblt} improves decoding robustness by transmitting redundant coded IBLTs, at the cost of increased communication. \riblt~\cite{lei2024riblt} eliminates parameter tuning by streaming coded symbols until decoding succeeds,  introducing high round complexity. CertainSync~\cite{keniagin2025certainsync} provides deterministic guarantees. In this work, we introduce \adaptiveiblt, which adjusts transmission size based on decoding feedback, balancing robustness and round efficiency without requiring prior knowledge of $d$.

Reconciliation has also been explored in high-throughput blockchain subsystems, particularly for transaction relay and mempool reconciliation~\cite{boskov2024out, ozisikgraphene2019}. 
Practical systems such as Shrec~\cite{han2020shrec} highlight that sketches with strong theoretical guarantees may become impractical once computation cost and transport-layer behavior are taken into account. 


In this work, we build upon the GenSync~\cite{bovskov2022gensync}, a standardized reconciliation framework, to evaluate modern reconciliation protocols under realistic network conditions and further implement \adaptiveiblt.

\section{Observations of LN Gossip Inefficiency}
\label{sec:ln-observation}

We conduct an empirical study of LN gossip behavior under typical operating conditions. We seek, seeking to quantify the magnitude of communication overhead during routine gossip.



We keep a standard Lightning Network node (default configuration) continuously connected to the public LN network. Each 10-minute run maintains active peers and normal gossip exchange while capturing TCP traffic and tracking growth of the local \texttt{gossip\_store}. Across runs, the \texttt{gossip\_store} grows at $\sim 500–1000$~bytes/s from new channel and node data, although transport-layer traffic is much higher.


On average, only about \emph{30\%} of the TCP-level traffic (including both payload and headers) directly contributes to persistent \texttt{gossip\_store} growth. The remaining \emph{70\%} consists of protocol overhead, including redundant gossip messages, control messages, timestamp-based filtering, and traffic required to maintain channel and node liveness. While absolute traffic volumes vary slightly across runs due to peer selection and network dynamics, the ratio between useful state growth and total transmitted data remains stable.

This persistent gap suggests that LN gossip inefficiency is a structural consequence of the flooding-based dissemination strategy. Even when peers already share highly overlapping gossip state, the protocol lacks a mechanism to exchange only differences. This motivates the exploration of reconciliation approaches that directly target difference exchange, rather than attempting to optimize message-level broadcast alone.


\section{Adaptive-IBLT}
\label{sec:adaptive}

\adaptiveiblt generalizes the IBLT protocol with a runtime feedback loop through which the sender adjusts future IBLT transmission size based on decoding progress. In so doing, this design eliminates the requirement of an accurate estimate of the number of differences between peers.

Our protocol starts with standard IBLT-based reconciliation using an initial number of cells (default $0.5\%$ of the set size). If decoding fails, the receiver notifies the sender, which doubles the cell count (default factor $2$) and resends the encoding. This repeats until decoding succeeds.

We extend \adaptiveiblt with \emph{partial decoding} to reduce communication and improve decoding success in marginal failure cases. When IBLT decoding fails, some elements yet be recovered; we cache them and exclude them from later rounds, lowering cost. Experiments (Section~\ref{sec:evaluation}) show reduced communication, especially near the decoding threshold, and improved latency when differences are hard to estimate.


\section{Evaluation of Sync Protocols}
\label{sec:evaluation}

We implemented \adaptiveiblt within the \texttt{GenSync} benchmarking framework~\cite{bovskov2022gensync}, together with a number of additional state-of-the-art reconciliation protocols.  
This unified framework allows us to compare the performance of the various data reconciliation protocols under the practical constraints. Our evaluation measures total time to reconcile (TTR) and communication cost across various workloads. \emph{We conclude that no protocol is globally optimal.}

Each reconciliation session runs between two local processes communicating over TCP sockets. We emulate realistic network conditions using the \texttt{mininet}~\cite{lantz2010mininet} framework, which enables packet-level control of latency, bandwidth, and loss. Our experiments use 10 ms round-trip latency, and bandwidth caps of 18~Mbps, all with 0.1\% packet loss (as in ~\cite{bovskov2022gensync}), and each experiment is repeated 100 times on independently sampled sets of 32-byte strings with 95\% confidence intervals.

Experiments were on an Intel Core i5-8400 CPU @ 2.80GHz and 16GB RAM, running Ubuntu 20.04 LTS. 
The \texttt{GenSync} framework is compiled with \texttt{g++ 11.4.0} \texttt{CMake 3.27}, \texttt{GMP 6.1.2}, and \texttt{NTL 10.5.0}. 
Protocol configurations and implementations are available at~\cite{gensync_repo}.

For each reconciliation run, we collect: \emph{Total Time to Reconcile} (TTR) - The wall-clock time from protocol initiation to successful reconciliation; \emph{Communication Cost} - The total number of bytes exchanged by both parties over the full reconciliation process, including all data and control.

\subsection{Total Reconciliation Time}
\label{sec:ttr}

\begin{figure}[t]
        \includegraphics[width=0.97\linewidth]{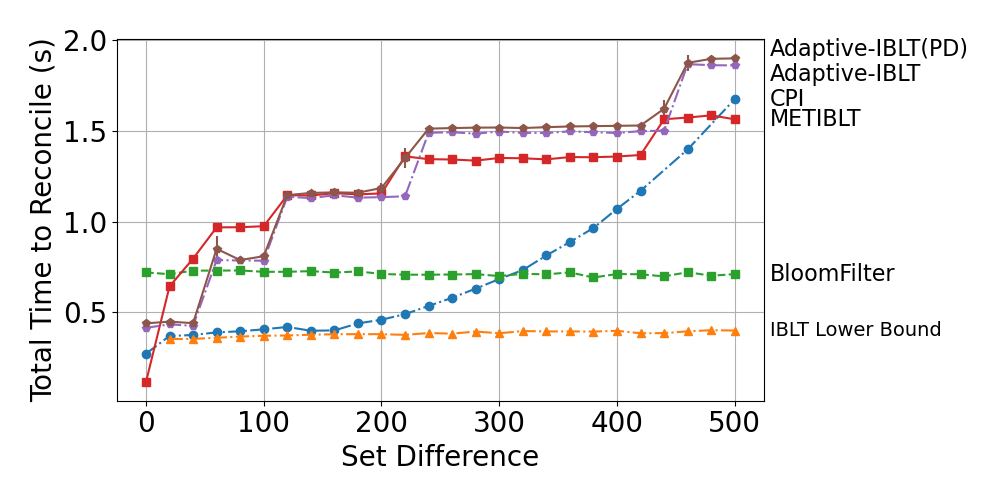}
    \caption{Total Time to Reconcile for some protocols.}
    \label{fig:all-syncs-ttr}
    \vspace*{-3ex}
\end{figure}

Figure~\ref{fig:all-syncs-ttr} compares reconciliation protocols under 10,ms RTT, 18,Mbps bandwidth, and 0.1\% packet loss. BloomFilter and IBLT (a ``lower bound'' assuming knowledge of $d$) maintain low TTR, with IBLT achieving the lowest latency. \metiblt has higher TTR at small $d$ due to redundant coded transmissions but scales more gracefully as $d$ increases. In contrast, \adaptiveiblt may incur higher TTR at large $d$ when doubling overshoots the true difference. 
Among interactive protocols (not shown), \interactive yields moderate TTR (20 seconds at $d=500$) for small differences but exhibits sharp spikes (and high variance), while \riblt reaches up to 60 seconds TTR, partly due to high round complexity from sending one cell per round.

\subsection{Communication Cost}
\label{sec:cc}

\begin{figure}[t]
    \centering
    \includegraphics[width=0.9\linewidth]{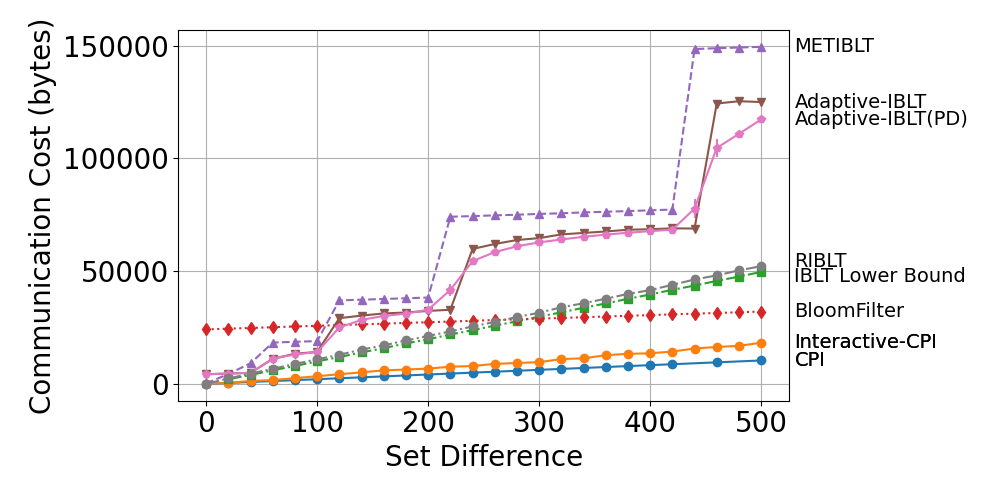}
    \caption{Communication Cost Comparison (Set Size = 10,000).}
    \label{fig:all-syncs-cc}
\end{figure}

Figure~\ref{fig:all-syncs-cc} presents total communication cost versus $d$.  CPI and \interactive achieve the lowest communication overhead, with BloomFilter slightly higher and IBLT scaling linearly.  CPI, BloomFilter, and IBLT 
require a known $d$.

\adaptiveiblt incurs higher cost due to geometric retries when $d$ is underestimated, while its partial decoding variant reduces this overhead by avoiding retransmission of already recovered elements. \metiblt has the highest communication cost, as its redundancy is independent of decoding success. \riblt achieves near-linear cost in $d$ with minimal redundancy, but transmits over many rounds.

\section{Implementation in Lightning Network}
\label{sec:ln-impl}

Lightning Network implementations rely heavily on timely broadcast of gossip messages to maintain an up-to-date view of the network topology. We suggest a more efficient design: performing a pre-connection gossip-state reconciliation using Gensync. In this architecture, a node executes a reconciliation protocol with trusted peers and can then proceed with standard Lightning connections with nearly identical gossip state. This avoids the large initial gossip flood required to “catch up”.


We run a vanilla Lightning node connected to a highly connected peer (\texttt{1ML.com node ALPHA}) for 1~hour to populate its gossip store, then take a snapshot. A second snapshot is taken $\sim$20 minutes later, reflecting 19 minutes 35 seconds of gossip-based flooding. The snapshots are 22.26,MB (77,862 entries) and 23.61,MB (82,068 entries). We run each set reconciliation protocol 20 times on these snapshots and measure synchronization time relative to the flooding time.

As shown in Figure 5, IBLTSync, \adaptiveiblt, \riblt, \metiblt, and \interactive all outperform the existing gossip-based reconciliation with respect to TTR, as they all take less than 19 minutes to complete successfully. CPISync (not shown in the figure) takes approximately 40 minutes and is the only protocol to perform worse than the gossip dissemination, due to the long computation time.

We also compare the communication costs of each reconciliation protocol.
Generally, IBLT-based protocols have much higher communication costs: \metiblt has the highest at 28 MB; \adaptiveiblt and \riblt are slightly lower at 22 MB and 18 MB repectively; IBLT serves as the baseline of 10 MB;. The CPI and \interactive protocols have the lowest communication costs at 1.5 and 1.8 MB respectively.

\begin{figure}[t]
    \centering
    \includegraphics[width=0.9\linewidth]{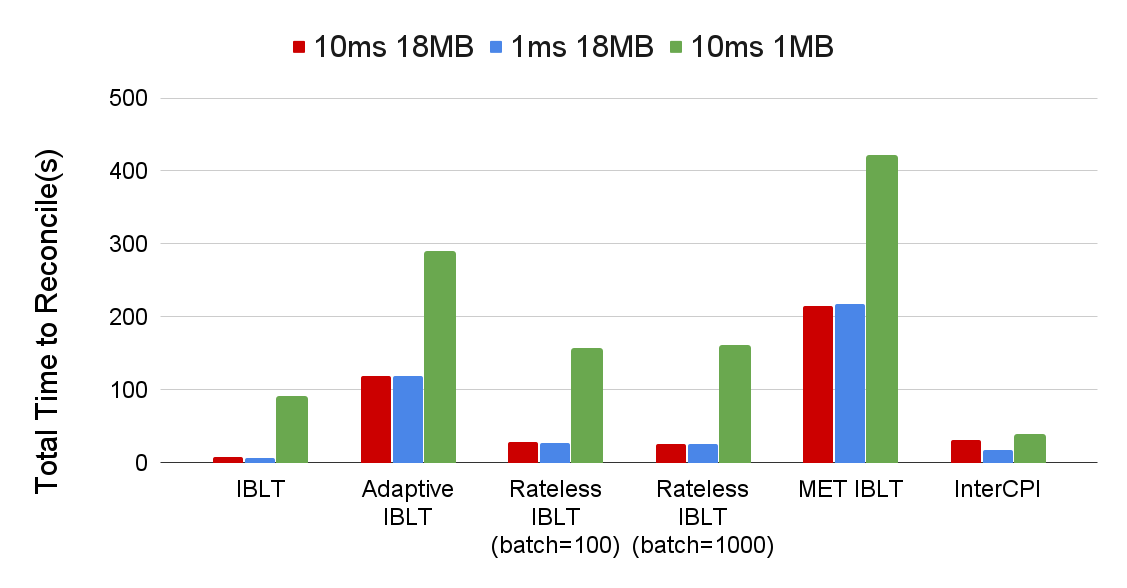}
    \caption{TTR comparison of reconciliation protocols across varying network bandwidths, evaluated on two gossip snapshots collected 20 minutes apart.}
    \label{fig:gossip_ttr}
\end{figure}


\section{Conclusion}
\label{sec:conclusion}

This paper revisits Lightning Network gossip through the lens of set reconciliation. Our packet-level measurements of flooding-based gossip showcase a persistent inefficiency: although the rate of growth of new data is fairly low, we observed significant amounts of low-level network traffic, suggesting structural redundancy in the current approach.


Motivated by these observations, we conduct a systematic, fair evaluation of reconciliation protocols in GenSync under realistic network conditions, focusing on those without prior knowledge of the set difference. Our results show protocol choice depends heavily on network constraints: redundancy-heavy designs (\metiblt) trade bandwidth for robustness, while highly interactive designs (\riblt and other multi-round protocols) suffer latency amplification under nontrivial RTTs. These findings reinforce that asymptotic communication complexity is insufficient to predict real performance.

To this point, we introduced \adaptiveiblt, a simple runtime-adaptive IBLT-based protocol, enhanced by partial decoding to reuse intermediate progress across rounds. 
Empirically, \adaptiveiblt provides a robust performance trade-off across diverse workloads and network settings. 
By evaluating on real gossip snapshots, we showed that reconciliation can reduce gossip convergence time by orders of magnitude.

Future work may include integrating additional protocols such as CertainSync~\cite{keniagin2025certainsync} into the evaluation framework in order to provide a broader comparison across rateless and deterministic designs.  We also envision studying reconciliation under more diverse churn and adversarial conditions, and exploring more designs to provide both fast startup and peer-to-peer freshness without relying on trusted providers.

\section*{Acknowledgments}
This work was supported in part by the National Science Foundation under Grant CNS-2210029 and by the Red Hat Collaboratory at Boston University.

\bibliographystyle{IEEEtran}
\bibliography{common, additional}

\end{document}

%% file: macros.tex
\newcommand{\junk}[1]{}
\newcommand{\BfPara}[1]{{\noindent {\bf #1.}}}

\newsavebox{\ieeealgbox}

\setlength{\parskip}{3pt}

\renewcommand{\thefigure}{\arabic{figure}}
\renewcommand{\theequation}{\arabic{equation}}
\renewcommand{\thetable}{\arabic{table}}

\newcommand{\ignore}[1]{}

\newcommand{\jnote}[1]{\medskip\noindent\begin{center}\fbox{\parbox{.9\textwidth}{{\bf Note:} \emph{#1} --- {\bf Jon}}}\end{center} \medskip}

\newcommand{\argmin}{\mathop{\mathrm{arg\,min}}}
\newcommand{\dmin}{\ensuremath{d_{\textrm{min}}}}
\newcommand{\Proof}{\textbf{Proof:\ }}
\newcommand{\floor}[1]{\lfloor #1 \rfloor}
\newcommand{\ceil}[1]{\lceil #1 \rceil}
\newcommand{\abs}[1]{|#1|}
\newcommand{\F}{{\mathbb{F}}}
\newcommand{\sym}[1]{\textup{\textsf{#1}}}
\newcommand{\aprime}{\ensuremath{\mathfrak{p}}}
\newcommand{\matchprob}{\rho_{\textsf{match}}}
\newcommand{\Itrans}{\ensuremath{I_{\textrm{trans}}}}
\newcommand{\polyone}[1]{\raisebox{0.15em}{$\chi$}_{\mbox{\footnotesize $#1$}}}
\newcommand{\poly}[2]{\raisebox{0.15em}{$\chi$}_{\mbox{\footnotesize $#1$}}(#2)} 
\newcommand{\dpoly}[3]{\poly{#1/#2}{#3}}
\def\nospaceqed{\vspace{6pt}\hfill \hbox{\vrule width4pt depth2pt height6pt}}
\def\qed{\nospaceqed\\}
\newcommand{\tagg}[1]{\textup{\texttt{#1}}}
\newcommand{\apriori}{\emph{a priori}}
\newcommand{\ie}{\emph{i.e., }}
\newcommand{\eg}{\emph{e.g. }}
\renewcommand{\paragraph}[1]{\medskip\noindent{\bf #1.}}
\newcommand{\sectref}[1]{Section~\ref{#1}}
\newcommand{\boundm}{{\ensuremath{\overline{m}}}}
\newcommand{\boundf}{{\ensuremath{\overline{f}}}}
\newcommand{\kw}[1]{\texttt{\textrm{#1}}}
\newcommand{\defeq}{\overset{\mathit{def}}{=}}
\renewcommand{\gcd}{\text{GCD}}
\newcommand{\degree}{\ensuremath{\mathit{degree}}}
\newcommand{\C}{\ensuremath{\mathbb{C}}}
\newcommand{\D}{\ensuremath{\mathbb{D}}}
\newcommand{\zero}{{\mathbf 0}}
\newcommand{\wt}{{\mathrm w}{\mathrm t}}
\newcommand{\Bullet}{({\large$\bullet$})}
\newcommand{\Star}{({\large\textbf{$\star$}})}
\newcommand{\dave}{\ensuremath{d_{\mbox{ave}}}}
\newcommand{\itembf}[1]{\item\ \textbf{#1}}
\newcommand{\al}{\fbox{what is \\al?}}
\renewcommand{\th}{\fbox{what is \\th?}}
\newcommand{\cn}{\fbox{what is \\cn?}}
\newcommand{\snr}{\fbox{what is \\snr?}}
\newcommand{\beq}{\begin{equation}}
\newcommand{\eeq}{\end{equation}}
\newcommand{\proofover}{\qed}
\newcommand{\eH}{\ensuremath{\mathcal{H}}}

\def\inline#1:{\par\vskip 7pt\noindent{\bf #1:}\hskip 10pt}
\def\qed{{\unskip\nobreak\hfil\penalty50%
  \hskip2em\hbox{}\nobreak\hfil$\square$%
  \parfillskip=0pt\finalhyphendemerits=0\par}}
\long\def\comment #1\commentend{}
\long\def\commabs #1\commabsend{}
\long\def\commful #1\commfulend{#1}

\newcommand{\mb}{\mathbf}
\newcommand{\bsy}{\boldsymbol}
\newcommand{\mbb}{\mathbb}
\newcommand{\ra}{\rightarrow}

\newcommand{\by}{\mb{y}}
\newcommand{\bY}{\mb{Y}}
\newcommand{\scrB}{\mathcal{B}}
\newcommand{\scrL}{\mathcal{L}}
\newcommand{\scrS}{\mathcal{S}}
\newcommand{\scrY}{\mathcal{Y}}
\def\liminf{\mathop{\rm lim\ inf}\limits}
\def\limsup{\mathop{\rm lim\ sup}\limits}%

\newcommand{\ZPETRAKE}{TREKS}
\newcommand{\rob}[1]{\left( #1 \right)}
\newcommand{\sqb}[1]{\left[ #1 \right]}
\newcommand{\cub}[1]{\left\{ #1 \right\}}
\newcommand{\veb}[1]{\left| #1 \right|}
\newcommand{\zo}{\cub{0,1}}

\newcommand{\enc}{\emph{ENC}}

\newcommand{\MEM}{\texttt{TAPE} }
\newcommand{\ACC}{\texttt{ACC} }
\newcommand{\subleqm}{\texttt{SUBLEQ\_M} }
\newcommand{\subleqALT}{\texttt{SUBLEQ\_\mbox{ALT}} }

\newtheorem{defn}{Definition}[section]
\newtheorem{lemma}{Lemma}[section]
\newtheorem{theorem}{Theorem}[section]
\newtheorem{cor}{Corollary}[theorem] 
\newtheorem{method}{Method}[section]
\newtheorem{alg}{Algorithm}[section]
\newtheorem{corsec}{Corollary}[section]     
\newtheorem{corlem}{Corollary}[lemma]       
\newtheorem{modif}{Modification}[theorem]   
\newtheorem{conjecture}{Conjecture}
\newtheorem{proposition}{Proposition}

\newcommand{\ddim}{\mathsf{ddim}}

\newcommand{\metiblt}{\textsc{METIBLT}\xspace}
\newcommand{\adaptiveiblt}{\textsc{AdaptiveIBLT}\xspace}
\newcommand{\riblt}{\textsc{RatelessIBLT}\xspace}
\newcommand{\interactive}{\textsc{InteractiveCPI}\xspace}

%% file: additional.bib
@String{Computing = "Computing" }

@inproceedings{minsky2002scalable,
  title={Scalable set reconciliation},
  author={Minsky, Yaron and Trachtenberg, Ari},
  booktitle={Proceedings of the Annual Allerton Conference on Communication, Control, and Computing},
  volume={40},
  number={3},
  pages={1608--1617},
  year={2002},
  organization={The University; 1998}
}

@article{minsky2003set,
  title={Set reconciliation with nearly optimal communication complexity},
  author={Minsky, Yaron and Trachtenberg, Ari and Zippel, Richard},
  journal={IEEE Transactions on Information Theory},
  volume={49},
  number={9},
  pages={2213--2218},
  year={2003},
  publisher={IEEE}
}

@article{bovskov2022gensync,
  title={Gensync: A new framework for benchmarking and optimizing reconciliation of data},
  author={Bo{\v{s}}kov, Novak and Trachtenberg, Ari and Starobinski, David},
  journal={IEEE Transactions on Network and Service Management},
  year={2022},
  publisher={IEEE}
}

@article{boskov2024out,
  title={Out-of-band transaction pool sync for large dynamic blockchain networks},
  author={Boskov, Novak and Chen, Xingyu and Simsek, Sevval and Trachtenberg, Ari and Starobinski, David},
  journal={International Journal of Network Management},
  volume={34},
  number={5},
  pages={e2265},
  year={2024}
}

@ARTICLE{lazaro2023metiblt,
  author={Lázaro, Francisco and Matuz, Balázs},
  journal={IEEE Transactions on Communications}, 
  title={A Rate-Compatible Solution to the Set Reconciliation Problem}, 
  year={2023},
  volume={71},
  number={10},
  pages={5769-5782},
  doi={10.1109/TCOMM.2023.3296630}}

@misc{keniagin2025certainsync,
      author = {Tomer Keniagin and Eitan Yaakobi and Ori Rottenstreich},
      title = {{CertainSync}: Rateless Set Reconciliation with Certainty},
      howpublished = {Cryptology {ePrint} Archive, Paper 2025/623},
      year = {2025},
      url = {https://eprint.iacr.org/2025/623}
}

@inproceedings{lei2024riblt,
author = {Yang, Lei and Gilad, Yossi and Alizadeh, Mohammad},
title = {Practical Rateless Set Reconciliation},
year = {2024},
isbn = {9798400706141},
publisher = {Association for Computing Machinery},
url = {https://doi.org/10.1145/3651890.3672219},
booktitle = {Proceedings of the ACM SIGCOMM 2024 Conference},
pages = {595–612},
numpages = {18},
series = {ACM SIGCOMM '24}
}

@inproceedings{ozisikgraphene2019,
author = {Ozisik, A. Pinar and Andresen, Gavin and Levine, Brian N. and Tapp, Darren and Bissias, George and Katkuri, Sunny},
title = {Graphene: efficient interactive set reconciliation applied to blockchain propagation},
year = {2019},
isbn = {9781450359566},
publisher = {Association for Computing Machinery},
doi = {10.1145/3341302.3342082},
booktitle = {Proceedings of the ACM Special Interest Group on Data Communication},
pages = {303–317},
numpages = {15},
location = {Beijing, China},
series = {SIGCOMM '19}
}

@inproceedings{han2020shrec,
author = {Han, Yilin and Li, Chenxing and Li, Peilun and Wu, Ming and Zhou, Dong and Long, Fan},
title = {Shrec: bandwidth-efficient transaction relay in high-throughput blockchain systems},
year = {2020},
isbn = {9781450381376},
publisher = {Association for Computing Machinery},
address = {New York, NY, USA},
url = {https://doi.org/10.1145/3419111.3421283},
doi = {10.1145/3419111.3421283},
booktitle = {Proceedings of the 11th ACM Symposium on Cloud Computing},
pages = {238–252},
numpages = {15},
location = {Virtual Event, USA},
series = {SoCC '20}
}

@inproceedings{lantz2010mininet,
author = {Lantz, Bob and Heller, Brandon and McKeown, Nick},
title = {A network in a laptop: rapid prototyping for software-defined networks},
year = {2010},
isbn = {9781450304092},
publisher = {Association for Computing Machinery},
address = {New York, NY, USA},
url = {https://doi.org/10.1145/1868447.1868466},
doi = {10.1145/1868447.1868466},
booktitle = {Proceedings of the 9th ACM SIGCOMM Workshop on Hot Topics in Networks},
articleno = {19},
numpages = {6},
location = {Monterey, California},
series = {Hotnets-IX}
}

@misc{gögge2022routingdelayLN,
      title={On the Routing Convergence Delay in the Lightning Network}, 
      author={Niklas Gögge and Elias Rohrer and Florian Tschorsch},
      year={2022},
      eprint={2205.12737},
      archivePrefix={arXiv},
      primaryClass={cs.NI},
      url={https://arxiv.org/abs/2205.12737}, 
}

@book{masteringLN,
  title        = {Mastering the Lightning Network},
  author       = {Antonopoulos, Andreas M. and Osuntokun, Olaoluwa and Pickhardt, Rene},
  year         = {2021},
  publisher    = {O'Reilly Media},
  address      = {Sebastopol, CA},
  note         = {Chapter 11: Gossip and the Channel Graph}
}

@misc{harvey2025gossipobserver,
  author       = {Harvey, Jon},
  title        = {Gossip Observer: Monitoring Gossip Propagation in the Lightning Network},
  year         = {2025},
  howpublished = {\url{https://github.com/jharveyb/gossip\_observer}},
}

@misc{ldkRGS,
  author       = {{Lightning Dev Kit Team}},
  title        = {Rapid Gossip Sync: Efficient Graph Synchronization for Lightning Nodes},
  year         = {2022},
  howpublished = {\url{https://lightningdevkit.org/blog/announcing-rapid-gossip-sync/}},
}

@INPROCEEDINGS{he2019improvedgossip,
  author={He, Xiaowei and Cui, Yiju and Jiang, Yunchao},
  booktitle={2019 International Conference on Cyber-Enabled Distributed Computing and Knowledge Discovery (CyberC)}, 
  title={An Improved Gossip Algorithm Based on Semi-Distributed Blockchain Network}, 
  year={2019},
  volume={},
  number={},
  pages={24-27},
  doi={10.1109/CyberC.2019.00014}}

@misc{gensync_repo,
  author       = {Xingyu Chen and others},
  title        = {GenSync: A Network-Aware Set Reconciliation Framework},
  year         = {2026},
  howpublished = {\url{https://github.com/nislab/gensync}},
}

@misc{lnsummit2024_gossip,
  author       = {{Lightning Network Developers}},
  title        = {LN Summit 2024 Notes: Summary and Commentary},
  year         = {2024},
  howpublished = {\url{https://delvingbitcoin.org/t/ln-summit-2024-notes-summary-commentary/1198}},
}


%% file: common.bib
@article{bloom70spacetime,
	Author = {B. H. Bloom},
	Journal = {Communications of the ACM},
	Month = {July},
	Number = 7,
	Pages = {422--426},
	Title = {Space/Time Trade-offs in Hash Coding with Allowable Errors},
	Volume = 13,
	Year = 1970}

@article{TMC03,
	Author = {D. Starobinski and A. Trachtenberg and S. Agarwal},
	Journal = {IEEE Trans. on Mobile Computing},
	Month = {January-March},
	Number = 1,
	Title = {Efficient {PDA} synchronization},
	Volume = 2,
	Year = 2003}

@inproceedings{goodrich2011invertible,
	Author = {Goodrich, Michael T and Mitzenmacher, Michael},
	Booktitle = {49th Annual Allerton Conference on Communication, Control, and Computing},
	Organization = {IEEE},
	Pages = {792--799},
	Title = {Invertible {B}loom lookup tables},
	Year = {2011}}
